# Thermodynamics of adsorption onto single-walled carbon nanotube homogeneous bundles

Fernando J.A.L. Cruz and José P.B. Mota*

*Requimte/CQFB, Departamento de Química, Faculdade de Ciências e Tecnologia, Universidade Nova de Lisboa, 2829-516 Caparica, Portugal. E-Mail: pmota@dq.fct.unl.pt*

The thermodynamics of adsorption onto multi-site adsorptive solids has been studied employing different probe molecules ($CH_4$, $C_2H_6$, $C_2H_4$, $C_3H_8$, $C_3H_6$), and conducting configurational-bias Grand Canonical Monte Carlo simulations. The phase space has been systematically accessed performing calculations at room temperature in the reduced pressure range $8.7 \times 10^{-9} < (p/p_0) < 0.9$. The solid consisted of homogeneous single-walled carbon nanotube bundles with an individual nanotube diameter distribution of $11.0 \text{ Å} \leq D \leq 18.1 \text{ Å}$. The resulting picture is interpreted in terms of the molecular nature of the adsorbate and the corresponding solid-fluid interactions. Results obtained indicate that confinement onto the bundles interior volume (interstitial and intratubular) is energetically more favorable than physisorption onto the bundles exterior volume (grooves and rounded surface), as indicated by the isosteric heat curves, $q_{st}$, as function of reduced pressure. Nonetheless, the zero-coverage properties suggest a crossover point to the former behavior, located at $D \approx 18 - 19 \text{ Å}$. Moreover, when interstitial confinement is not inhibited by geometrical considerations, it was possible to establish a general relative order for the isosteric heat of adsorption: $\left(q_{st}^0\right)^{interstitial} > \left(q_{st}^0\right)^{intratubular} > \left(q_{st}^0\right)^{grooves} > \left(q_{st}^0\right)^{surface}$.

## I. INTRODUCTION

Molecular confinement onto nanoscale pores is of fundamental and applied importance in a wide range of physical, chemical and biological processes.[1-7] As model nanopores of cylindrical geometry, single-walled carbon nanotubes (SWCNTs)[8,9] have been receiving a great amount of attention due to their unique and exciting features, such as optical[10] and electronic properties.[11] Amongst several other applications,[12] SWCNTs have been proposed to be used as building blocks in composites,[13] chemical sensors,[14] separating agents of organic vapors[15] and as storage nanomaterials for hydrogen[16] and methane.[17,18] Whatever the specific application is, in a great number of cases it involves the interaction of organic fluids with the solid lattice by means of physisorption confinement. Similarly to a graphene sheet, SWCNTs exhibit a π–electron cloud on their walls, arising from the $sp^2$ hybridization of carbons atoms. Due to the strong van der Waals interactions[1] resulting from this charge distribution, SWCNTs samples are usually obtained as a collection of individual tubes aggregated in the form of heterogeneous spaghetti-like structures called bundles.[19] As depicted in Figure 1 for the usual hexagonal lattice,[20-22] the available adsorption sites are distributed over the internal porous volume of the bundles (intratubular and interstitial adsorption) and on their peripheral exterior surface (rounded surface and groove sites). The existence of these distinct types of adsorption sites is a fundamental difference between SWCNTs and other carbon materials, such as activated carbon, that needs to be addressed in any accurate study of the adsorptive properties of these structured nanomaterials. For example, both the interstitials and groove sites can be interpreted as quasi linear arrays for the physical realization of matter in one dimension.[23,24] However, interstitial sites only become available for adsorption when individual nanotube diameter is large enough.[25] Moreover, both the intratubular and exterior rounded surface will be able to physisorb a molecule in a variety of different adsorptive sites, resulting in two or even three dimensional adsorption. Because these sites are intrinsically different, what will be the corresponding energetic differences between them, and how will those differences relate to measurable properties such as individual tube diameter and bulk fluid pressure (chemical potential)? These are some of the questions we'll address in the present study.

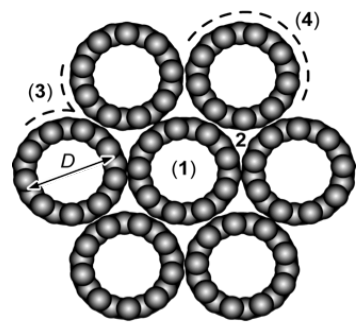

**Figure 1.** Different adsorption sites on an homogeneous bundle of SWCNTs with individual tube diameter $D$: (1) intratubular, (2) interstitial channel, (3) external groove, (4) external rounded surface. Sites 1 and 2 comprise the internal pore volume of the bundle, whereas sites 3 and 4 are both located on the exposed surface of the outermost nanotubes.

Due to its spherical geometry and well characterized physical properties, methane is one of the most commonly used probe molecules to study the thermodynamic properties of adsorption onto single-walled carbon nanotubes bundles.[18, 20, 23-31] Properties such as the isosteric heat of adsorption are used as the relevant parameters to monitor the mechanism of physisorption confinement. Kowalczyk et al have performed GCMC simulations of $CH_4$ adsorbed onto a (10,10) SWCNT homogeneous bundle at 293 K, considering only interstitial and intratubular adsorption.[18] They concluded that the isosteric heat of adsorption increased non-linearly with adsorbate loading up to a total maximum of 20.5 kJ/mol, and started decreasing henceforward. A similar bundle was employed by Jiang et al to study the adsorption of linear alkanes $C_n$ ($n \leq 5$) at 300 K,[32] and they have reported total adsorption zero–coverage isosteric heats, $q_{st}^0$, in the range 18.27 kJ/mol – 53.36 kJ/mol. Recently, Johnson and co-workers have addressed the adsorption of $CH_4$ on heterogeneous and homogeneous bundles, using both simulation techniques and low-temperature (77K) experimental measurements.[25] The isotherms and isosteric heats for the groove sites and outside surface of heterogeneous bundles were found to be very close to those for the homogeneous bundles, indicating that bundle heterogeneity is of little consequence on external surface adsorption. This is not surprising, bearing in mind that for a small molecule as methane, a bundle exterior surface can behave similarly to a planar graphite sheet, depending on the spacing between adjacent tubes. A similar premise has been postulated by Shi et al who have observed that groove sites located in either homogeneous or heterogeneous closed-ended bundles are indeed very similar.[28] A very interesting experimental study was conducted by Bienfait et al using closed-ended SWCNTs homogeneous bundles of 17 ± 1 Å diameter.[29] Although the intra-tubular volume was neglected, their studies clearly showed two preferential adsorptive sites on the sample, as evidenced by two plateaus in the isosteric heat curve as function of coverage. The grooves and interstitial sites seemed to exhibit roughly 62 % higher isosteric heats than the external rounded surface of the bundles. The adsorption simulations and neutron diffraction experiments with closed-ended homogeneous SWCNT bundles of Johnson and co-workers,[24] seem to corroborate the idea that interstitials and groove sites are energetically more favorable for adsorption than surface sites, and therefore are the ones initially becoming populated by $CH_4$ molecules during adsorption from a bulk gas onto a solid sample.

The aforementioned studies are starting contributions to the effort of building up a molecular-level picture of the phenomena governing adsorption onto multi-site solids. However, an unabridged understanding of the microscopic details involved is still very incomplete, particularly in the case of intratubular adsorption, whose mechanism is known to play a predominant role in the low pressure limit.[33] Amongst other issues that also need carefully investigation is the of the probe molecule itself, which can be either spherical or linear, as well as its spatial orientation[34]. Two of the simplest examples of these latter species are ethane ($C_2H_6$) and ethylene ($C_2H_4$), whose main difference between the molecules lies in the existence of an unsaturated π bond in ethylene; this kind of unsaturated chemical bond attributes the molecule a certain degree of rigidity and at the same time makes it smaller compared with the saturated analogue. How will the structure of the probe molecule and its chemical nature affect the thermodynamical properties of adsorption? What is the relationship between molecular length and isosteric heat of adsorption, expressed in terms of the number of carbon atoms located on the molecular skeleton? It is expected that some of the interstitial channels will not be physically available to all adsorbents, essentially due to: *i)* individual diameter of a single SWCNT in a bundle, and *ii)* adsorbent molecular diameter.

We have already presented a detailed structural analysis of commercially available SWCNT bundles,[19, 35] which was been validated by comparison with Raman scattering and experimental adsorption data of several organic fluids,[36, 37] nitrogen[19] and water.[38] In the present work, we address the particular case of an ideal open-ended SWCNT sample comprising homogeneous bundles. Experimentally, samples are usually prepared as heterogeneous nanopores, with some ends blocked, possessing a distribution of individual tube diameters[35] located in the range 11.0 Å < D < 18.1 Å. However, it is possible to sort out the individual tubes according to their diameter as well as treating the sample with physico-chemical processes to unblock the closed ends.[39-42] We will systematically conducting Grand Canonical Monte Carlo (GCMC) simulations[43-45] of $CH_4$, $C_2H_6$, $C_2H_4$, $C_3H_8$, and $C_3H_6$, at room temperature, onto different adsorption sites of the bundle (both inter- and intratubular volume, grooves, and external surface). The pressures explored in the molecular simulations (1 × $10^{-7}$ bar < $p$ < 105 bar) are within a relative pressure range of $8.67 \times 10^{-9} < (p/p_0) < 0.9$, where $p_s$ is the saturation pressure of the vapor adsorbate at the corresponding bulk fluid temperature. By eliminating impurities, polydispersity in nanotube diameter, and pore blockage, we will be addressing the most favorable bundle structure for application of SWCNTs as membranes, molecular sieves, and gas storage media.

The remaining of the paper is organized as follows. In the next section, the molecular simulation force-field and methodology will be presented along with the model employed for the SWCNTs bundle. In Section III, the calculated isosteric heats of adsorption and relative population curves are recorded and interpreted in terms of molecular nature of the probe molecules and the bundle structural characteristics. Extrapolation of data towards the limit of very low pressure will allow the determination of zero-coverage isosteric heats, and the corresponding study of the solid-fluid interactions. Whenever experimental data are available, comparisons will be drawn. Finally, some remarks and future guidelines will conclude.

## II. MOLECULAR MODEL AND SIMULATION DETAILS

The force field adopted for the five adsorbates, methane [$CH_4(sp^3)$], ethane [$CH_3(sp^3)$–$CH_3(sp^3)$], ethylene [$CH_2(sp^2)$=$CH_2(sp^2)$], propane [$CH_3(sp^3)$–$CH_2(sp^3)$–$CH_3(sp^3)$], and propylene [$CH_2(sp^2)$=$CH(sp^2)$–$CH_3(sp^3)$] — is the transferable potential for phase equilibria (TraPPE).[46, 47] This force field is based on an united-atom (UA) model where the $CH_4(sp^3)$, $CH_3(sp^3)$, $CH_2(sp^2)$, $CH_2(sp^3)$, and $CH(sp^2)$ groups are treated as single interaction sites. The nonbonded interactions between pseudo-atoms on different adsorbate molecules, as well as the interactions between carbon atoms of a nanotube[48-50] and pseudo-atoms of adsorbate molecules, are governed by the Lennard–Jones (LJ) 12–6 potential,

$$u(r_{ij}) = 4\varepsilon_{ij}\left[(\sigma_{ij}/r_{ij})^{12} - (\sigma_{ij}/r_{ij})^6\right], \quad (1)$$

where $r_{ij}$ is the intermolecular distance between sites $i$ and $j$. The potential well depths, $\varepsilon_i / k_B$ ($k_B$ is the Boltzmann constant), and collision diameters, $\sigma_i$, are listed in Table I. The cross terms are obtained using the classical Lorenz–Berthelot combining rules,[44, 51] $\varepsilon_{ij} = \sqrt{\varepsilon_i \varepsilon_j}$ and $\sigma_{ij} = (\sigma_i + \sigma_j)/2$. A spherical potential truncation for pairs of pseudo-atoms separated by more than 14 Å is enforced,[46] and analytical tail corrections are not applied.

**Table I:** Lennard–Jones parameters for the fluid TraPPe-UA force field[46, 47] and for the SWCNT atoms.[48-50]

| Pseudo-atom | $\varepsilon_i / k_B$ (K) | $\sigma_i$ (Å) |
|---|---|---|
| C (SWCNT) | 28.0 | 3.400 |
| $CH_4$ (methane) | 148.0 | 3.730 |
| $CH_3$ ($sp^3$) | 98.0 | 3.750 |
| $CH_2$ ($sp^3$) | 46.0 | 3.950 |
| $CH_2$ ($sp^2$) | 85.0 | 3.675 |
| $CH$ ($sp^2$) | 47.0 | 3.730 |

In the TraPPE-UA force field all bond lengths are fixed; the length of the $CH_x$–$CH_y$ bond is 1.54 Å, whereas that of the $CH_x$=$CH_y$ bond is 1.33 Å. The harmonic bond-bending potential, $u_{bend}(\theta)$, along the three pseudo-atoms of either propane or propylene is given by $u_{bend} = k_\theta (\theta - \theta_0)^2 / 2$. For propane, the force constant is $k_\theta / k_B$ = 62500 K rad$^{-2}$ and the equilibrium bending angle is $\theta_0 = 114.0°$; the corresponding values for propylene are $k_\theta / k_B$ = 70420 K rad$^{-2}$ and $\theta_0 = 119.7°$.

At ambient temperature the SWCNTs can be reasonably approximated as smooth structureless nanocylinders. For a pseudo-atom of an adsorbate molecule located at a nearest distance $\delta$ from the central axis of a nanotube, an effective potential, $U_{sf}(\delta)$, is developed by integrating the LJ solid–fluid potential, $u_{sf}(r)$, over the positions of all wall atoms of the nanotube (whose length is assumed to be infinite):

$$U_{sf}(\delta) = 4\rho_s R \int_0^\pi u_{sf}(r) d\theta dz,$$
$$r^2 = R^2 + \delta^2 - 2\delta R \cos\theta, \quad (2)$$

Where $R = D / 2$ is the pore radius, $z$ is the distance along the cylinder axis, $\theta$ is the radial angle, and $\rho_s = 0.382$ Å$^2$ is the atomic surface density of the SWCNT wall. By integrating over $z$ and $\theta$, eqn. 2 is reduced to a 1–D potential that is a function of $\delta$ only:

$$U_{sf}(\delta) = \pi^2 \rho_s \varepsilon_{sf} \sigma_{sf}^2 \left\{ \frac{63}{32}\left[\frac{R-\delta}{\sigma_{sf}}\left(1+\frac{\delta}{R}\right)\right]^{-10} \Phi(9,\delta/R) - 3\left[\frac{R-\delta}{\sigma_{sf}}\left(1+\frac{\delta}{R}\right)\right]^{-4} \Phi\left(3,\delta/R\right) \right\} \quad (3a)$$

$$\Phi(\alpha,\beta) = F(-\alpha/2, -\alpha/2, 1; \beta^2) \quad (3b)$$

Where $F(\alpha,\beta,\gamma;\delta)$ is the hypergeometric function. To speedup the calculation of $U_{sf}(\delta)$, eqn. 3a is tabulated on a grid with 31 knots equally spaced in $\delta^2$. During the simulations $U_{sf}(\delta)$ is reconstructed from the tabulated information using cubic Hermite polynomial interpolation.

Figure 2 shows the cross sections of the unit simulation boxes employed to study adsorption onto the bundle internal and external volume. The simulation boxes depicted in Figs. 2a and 2b are for intrabundle adsorption, whereas that shown in Figure 2c is for adsorption on the exterior volume of a bundle. It is worth noting that only the gray areas represent effective volume probed during the simulation; thus, the simulation box of Figure 2a comprises the internal volume of a cylinder, that in Figure 2b is a parallelepiped, and the one in Figure 2c is obtained by subtracting one quarter of a cylinder from the two bottom edges of a parallelepiped. The nanotubes in Figure 2b are arranged in the usual close-packed hxagonal lattice. The intertubular distance for all simulations is kept fixed at 3.4 Å to mimic SWCNTs adhering to each other via van der Waals forces forming bundles. The faces of each simulation box implement periodic boundary conditions, except for the top face of the box in Figure 2c, which is a reflecting wall, and the bottom face of the same box, which is blocked by the outermost shell of nanotubes in the bundle. The actual length of each simulation box is a function of the imposed adsorptive pressure.

To calculate the solid–fluid interaction potential of a pseudo-atom located inside a nanotube within a bundle, it suffices to sum the interactions of the pseudo-atom with the confining tube and the six nearest neighbors. The corrugation effect of the neighboring tubes is very small and, for practical purposes, does not affect the cylindrical symmetry of the total interaction potential. Therefore, when only intratubular adsorption is of interest it is computationally more efficient to employ the cylindrical simulation box shown in Figure 2a, which consists of the intratubular volume of a single nanotube under a force field that includes the additional contribution from the six nearest neighboring tubes. The simulation box shown in Figure 2b is employed to study the combined effect of intratubular adsorption and adsorption in the interstitial channels (where three tubes meet). An interstitial region can be divided into three symmetric volumes with pentagonal-like cross sections (some of the edges are denoted by the dotted lines). The solid–fluid potential for a pseudo-atom in one of those volumes is calculated by summing the interactions of the pseudo-atom and the four nearest tubes, as indicated in Figure 2b by the dark arrows. We have observed that including farther nanotubes does not significantly change the interaction potential.[52] Figure 2c shows the cross section of the unit simulation box for GCMC study of adsorption onto the exterior volume of a bundle. It has been shown previously[52] that to determine the overall interaction potential between a pseudo-atom of a sorbate molecule and the peripheral surface of the bundle it suffices to consider the interactions between the pseudo-atom and its five nearest nanotubes (three on the outermost shell and two on the second shell). Including farther nanotubes has a minimum impact on the total solid-fluid interaction potential. Notice that the nanotubes are not part of the simulation box itself and, therefore, molecules are not allowed to adsorb inside of them.

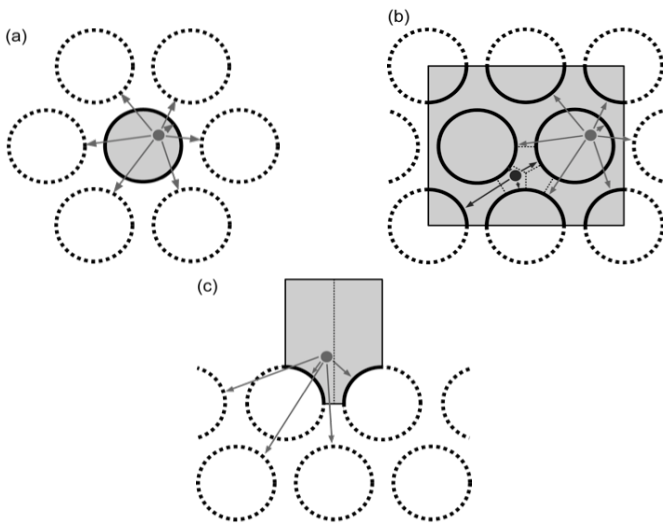

**Figure 2.** Cross sections of the unit simulation boxes for GCMC simulation of fluid adsorption onto different adsorption sites of an homogeneous bundle of open-ended SWNTs: *a)* intratubular volume; *b)* intratubular and interstitial channels; *(c)* exterior volume. The gray area represents the effective volume probed during the simulations.

To enhance the sampling of configurational space and increase the acceptance rate of the molecule insertion or removal step for the largest adsorbates (propane and propylene), we resort to configurational-bias sampling techniques.[53-56] In the configurational-bias method a flexible molecule is grown atom-by-atom towards energetically favorable conformations, leading to a scheme which is orders of magnitude more efficient than the traditional method of random growth. For the placement of the first pseudo-atom of an adsorbate molecule, $k_1 = 10$ random positions in the simulation box are generated, and one is selected with a probability $\exp(-\beta U_{1,i}^{ext})/\sum_j \exp(-\beta U_{1,i}^{ext})$, where $U_{1,i}^{ext}$ is the external energy of the pseudo-atom at the $j^{th}$ trial position interacting with the nanotubes and with the pseudo-atoms of the other adsorbate molecules. For each of the remaining two pseudo-atoms ($m = 2,3$) of the molecule, $k_m = 5$ trial positions are generated with a probability proportional to $\exp(-\beta U_{1,i}^{ext})$. These positions are distributed on the surface of a sphere centered on the previously inserted pseudo-atom of the molecule and whose radius is equal to the bond length. Each set of $k_m$ trial orientations is generated using the internal part of the potential $U_{m,j}^{int}$, whose probability depends on which type of pseudo-atom is being inserted; for the second atom ($m = 2$) the internal potential energy is zero and, as a result, the trial positions are randomly distributed on a sphere; for the third pseudo-atom ($m = 3$) the internal potential energy includes bond bending. For ($m = 3$) the trial positions are distributed on the edge of the disk which forms the base of a cone with apex at the center of the previously inserted bead and slant height equal to the bond length. For each trial position $j$ ($j = 1,...,k_m$) the external energy $U_{m,j}^{ext}$ is calculated for interaction with the nanotubes, with the pseudo-atoms of the other adsorbate molecules. From among the $k_m$ trial positions, one is selected with a probability $\exp(-\beta U_{m,i}^{ext})/\sum_j \exp(-\beta U_{m,j}^{ext})$. During this growth process a bias is introduced, but is removed by adjusting the acceptance rules.

Besides the usual trial step of molecule insertion/deletion, where the acceptance rate is enhanced by resorting to configurational-bias techniques, three additional types of Monte Carlo (MC) moves involving only individual molecules are necessary to sample the internal configuration of the simulation box: translation, rotation about the center-of-mass, and configurational-bias partial regrowth to change the internal conformation of a molecule. Each run is equilibrated for at least $2 \times 10^4$ Monte Carlo cycles followed by at least an equal number of cycles for the production period. Each cycle consists of $0.8N$ attempts to translate a randomly selected molecule, $0.2N$ trial rotations, $0.2N$ attempts to change the conformation of a molecule using configurational-bias partial regrowth, and max (20,0.2$N$) molecule insertion/deletion steps. Here, $N$ is the number of molecules in the simulation box at the beginning of each cycle. The maximum displacement for translation and angle for rotation are adjusted during the equilibration phase to give a 50% acceptance rate. Standard deviations of the ensemble averages are computed by breaking the production run into five blocks.

## III. RESULTS AND DISCUSSION

We have carried out Grand Canonical Monte Carlo simulations at room temperature (303.2 K – 298.1 K) and spanning a reduced pressure range of $8.59 \times 10^{-7} < (p / p_0) < 0.9$, where $p_0$ is the saturation pressure of the fluid [ref?] at the simulation temperature. This pressure region was extended down to $(p / p_0) = 8.67 \times 10^{-9}$ for $C_3H_8$ and $C_3H_6$. The isosteric heat of adsorption, $q_{st}$, is related to the amount of heat released when a molecule gets adsorbed on a substrate. Theoretically, this quantity can be calculated based on the definition given by Nicholson and Parsonage[43] (eqn. 4), where brackets denote the ensemble average, $N$ is the number of particles in the system, $U$ is the system's configurational energy, $k_B$ is the Boltzmann constant, and $T$ is the absolute temperature.

$$q_{st} = \frac{\langle U \rangle \langle N \rangle - \langle UN \rangle}{\langle N^2 \rangle - \langle N \rangle \langle N \rangle} + k_B T \qquad (4)$$

The results thus obtained are recorded in Figure 3 as a function of reduced pressure. The dispersion of data for methane external adsorption is statistical, and does not influence the curves general trends. Note that the upper and lower group of curves represents the bundle interior and exterior adsorption, respectively. In order to better clarify the distribution of molecules between the four different available adsorptive sites (*cf.* Figure 1), we have calculated the ratios of groove and intratubular adsorbed molecules on the bundles exterior and interior volume, respectively. For that, we define the ratio of molecules on groove sites as $N^{(g,s)} / N^{(s)}$, where $N^{(g,s)}$ is the number of molecules on the grooves and $N^{(s)}$ is the total number of adsorbed molecules (grooves and external surface). Similarly, the ratio of molecules on intratubular sites is defined by $N^{(i,v)} / N^{(v)}$, where $N^{(i,v)}$ is the number of molecules inside the nanotubes, and $N^{(v)}$ is the total number of adsorbed molecules (intratubular and interstitials).

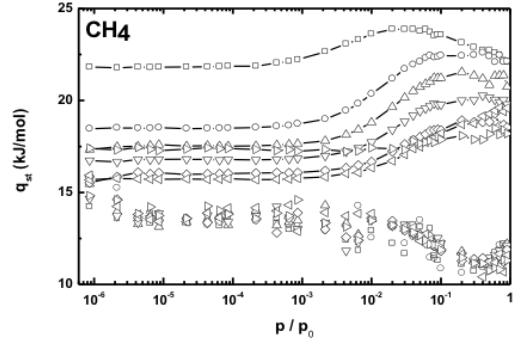
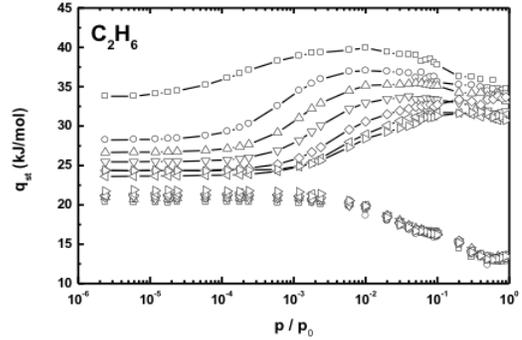
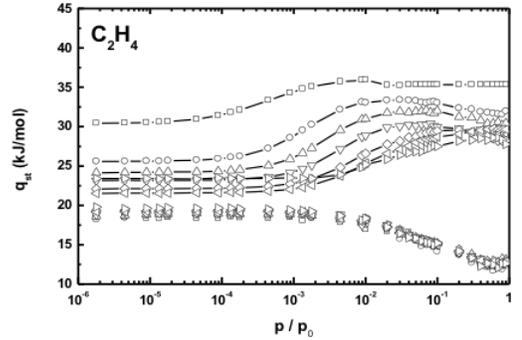
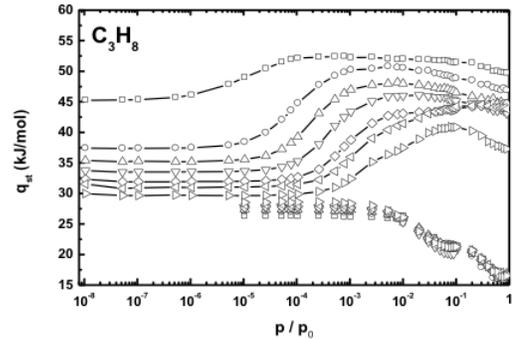

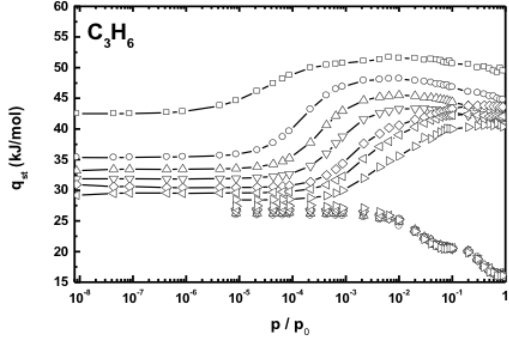

**Figure 3. Isosteric heat curves as a function of reduced pressure, ($p / p_0$),** on the bundle's interior (upper curves), intratubular and interstitial, and exterior (lower curves), grooves and rounded surface, adsorptive volumes. Note that ($p / p_0$) has been extended down to $8.67 \times 10^{-9}$ for $C_3H_8$ and $C_3H_8$ to validate their bundle interior adsorption general trends. Solid lines are guides to the eye. □ 11.0 Å, ○ 12.9 Å, △ 13.8 Å, ▽ 14.7 Å, ◇ 15.8 Å, ◁ 16.6 Å and ▷ 18.1 Å.

Before proceeding into a more detailed analysis of the results, some general remarks can already be established. Isosteric heats for interior adsorption are always larger than for external adsorption, and this difference becomes particularly relevant in the medium to high pressure region; for $p / p_0 > 10^{-2}$ these differences increase from (5 – 10 kJ/mol) to (10 – 25 kJ/mol) as the system approaches saturation conditions and molecular weight increases. It is clear from Figure 2 that isosteric heats are rather independent from individual tube diameter for adsorption on the bundles exterior volume; by contrast, $q_{st}$ values vary markedly with diameter when adsorption takes place inside the bundles. This latter trend gets attenuated as the fluids approach saturation, and the diameter becomes a less relevant factor. At high pressures entropy is the dominant thermodynamical variable, as opposite to energy under vacuum/low-pressure.

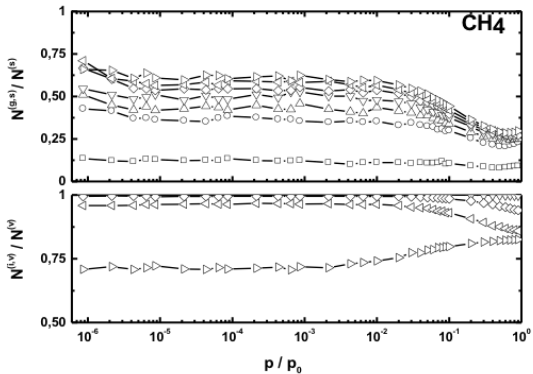

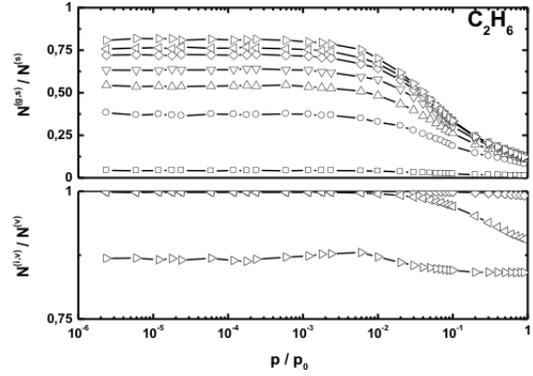

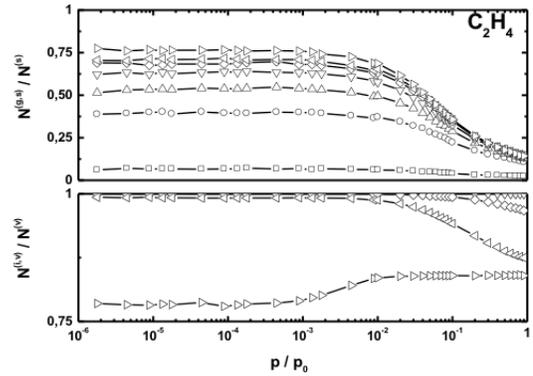

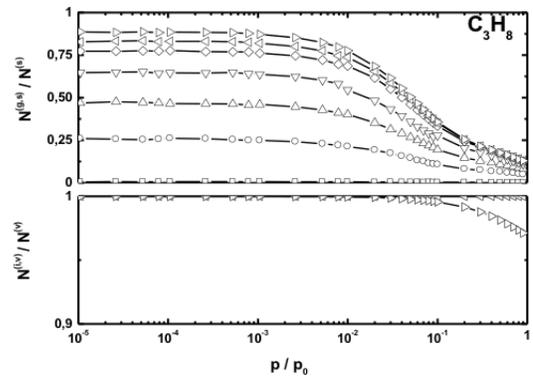

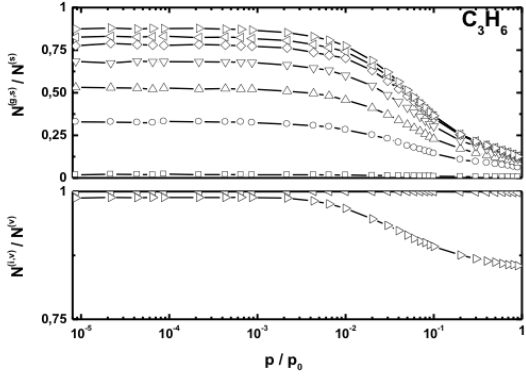

**Figure 4. Ratios of adsorbed molecules** between grooves and total exterior volume (grooves and surface), $N^{(g,s)} / N^{(s)}$, and between intratubular and total interior volume (intratubular and interstitials), $N^{(i,v)} / N^{(v)}$; $N^{(g,s)}$ and $N^{(i,v)}$ are, respectively, the number of molecules adsorbed on groove sites and inside the tubes (intratubular). Solid lines are guides to the eye. □ 11.0 Å, ○ 12.9 Å, △ 13.8 Å, ▽ 14.7 Å, ◇ 15.8 Å, ◁ 16.6 Å and ▷ 18.1 Å.

### A. Bundle interior volume: intratubular and interstitial channels

As shown in Figure 3, after an initial flat plateau the isosteric heat increases non-linearly with pressure, until reaching a maximum value characteristic of ideal SWCNT bundles, $q_{st}^{\max}$. Our result for CH$_4$ at a 13.8 Å diameter, with a maximum isosteric heat of $q_{st}^{\max}$ = 21.5 ± 0.3 kJ/mol, is in very good agreement with the previously reported value for an homogeneous (10,10) bundle ($D$ = 13.56 Å) of $q_{st}^{\max}$ = 20.5 kJ/mol.[18] This maximum results from the optimal interplay between the solid-fluid and fluid-fluid interaction energies. After that maximum, fluid-fluid interactions become more and more intense, caused by the condensation of fluids due to high pressure (chemical potential), and therefore the isosteric heat tends to decrease. The maximum isosteric heat values calculated are given in Table II for all the individual tube diameters probed. The results recorded in Table II suggest two general trends: *i)* $q_{st}^{\max}$ increases with the decrease of individual tube diameter, reflecting the enhanced solid-fluid interactions in the smaller nanotubes, and *ii)* $q_{st}^{\max}$ increases with the number of carbon atoms on the probe molecule, essentially due to the increase of interaction sites on the fluid.

**Table II.** Maximum isosteric heat of adsorption calculated for the bundle's interior volume, probed for all the individual tube diameters studied.

| | | | $q_{st}^{\max}$ | | |
|---|---|---|---|---|---|
| $D$ / Å | CH$_4$ | C$_2$H$_6$ | C$_2$H$_4$ | C$_3$H$_8$ | C$_3$H$_6$ |
| 11.0 | 23.9 ± 0.1 | 39.8 ± 0.2 | 35.9 ± 0.1 | 52.4 ± 0.4 | 51.7 ± 0.6 |
| 12.9 | 22.6 ± 0.3 | 37.0 ± 0.4 | 33.4 ± 0.3 | 50.8 ± 0.1 | 48.2 ± 0.1 |
| 13.8 | 21.5 ± 0.3 | 35.5 ± 0.4 | 32.1 ± 0.5 | 48.2 ± 0.4 | 45.5 ± 0.1 |
| 14.7 | 20.3 ± 0.4 | 33.8 ± 0.2 | 30.4 ± 0.3 | 46.1 ± 0.1 | 43.3 ± 0.3 |
| 15.8 | 19.3 ± 0.4 | 33.8 ± 0.6 | 29.6 ± 0.3 | 44.7 ± 0.1 | 43.8 ± 0.3 |
| 16.6 | 19.6 ± 0.4 | 31.7 ± 0.2 | 29.8 ± 0.6 | 44.5 ± 0.4 | 42.4 ± 0.2 |
| 18.1 | 18.5 ± 0.4 | 31.8 ± 0.5 | 28.6 ± 0.3 | 42.0 ± 0.4 | 41.0 ± 0.7 |

We can postulate a threshold of $N^{(i,v)} / N^{(v)}$ = 0.995 above which we consider that, within the statistical error of the calculations, all adsorbed molecules will be confined inside the tubes and interstitial adsorption can be safely neglected. By inspection of Figure 4 one can conclude that interstitial adsorption is always inhibited in the smaller nanotube diameters, 11.0 Å < $D$ < 14.7 Å. This has to do with geometrical limitations of the interstitial channels themselves, which are too narrow to physically accommodate fluid molecules. A different situation is observed at larger diameter pores, which form interstitials able to adsorb particles, at least under medium to high pressure conditions (15.8 Å and 16.6 Å). The largest diameter studied, 18.1 Å, exhibits an extreme situation, where significant interstitial adsorption can occur even in the very low pressure limit, $(p / p^0) < 10^{-3}$. It now becomes clear that the initial plateau in Figure 3 mainly corresponds to intratubular adsorption: it is constant while molecules are being adsorbed into a circular monolayer around the inner tube walls. A very interesting case is the phenomenon occurring at 11.0 Å, whose plateau is remarkably further away from the main group of curves. At this narrow diameter, not only molecules are not interstitially adsorbed, but they also fit perfectly well into that tube, lying with an orientation parallel to the tube main axis. As previously mentioned, the exception is the $D$ = 18.1 Å case, where molecules start being adsorbed both on the interstitial and intratubular sites, and whose corresponding $q_{st}$ curves lie above the lines for $D$ = 16.6 Å. This crossover happens for CH$_4$, C$_2$H$_6$ and C$_2$H$_4$, but not for C$_3$H$_8$ and C$_3$H$_6$. As an illustration to these phenomena, we have determined the concentration field gradient inside equilibrated simulation boxes, and represented it in the $x - y$ plane perpendicular to the nanotube main axis, $z$. For the sake of simplicity, results are only presented for methane, at two different diameters and reduced pressure, but representative of the main situations occurring in the whole phase and fluid space (Figure 5). Starting at the lowest diameter (Figure 5a), it is evident that interstitial adsorption is totally absent and intratubular confinement is the only relevant phenomenon. If we now move towards the higher diameter tube, equilibrated at the lowest reduced pressure (Figure 5b), one can observe that molecules initially get confined both in the interstitials and in the intratubular volume. As pressure increases, the intratubular monolayer approaches completeness and isosteric heat approaches its maximum. After reaching that maximum, $q_{st}$ starts to decrease due to the formation of an intratubular condensed phase lying parallel to the nanotube main axis (Figure 5c).

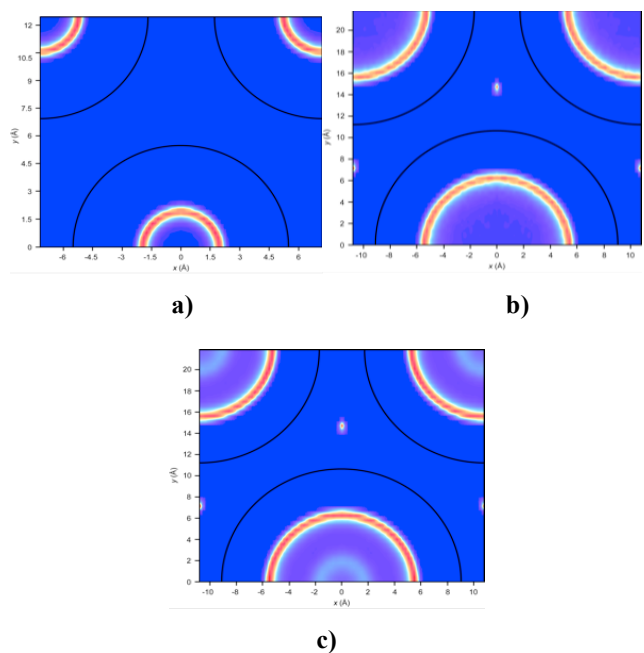

a)

b)

c)

**Figure 5. Contour plots of local concentration field for CH$_4$ on the $x - y$ plane, perpendicular to the nanotube main axis.** Data plotted for a bundle with individual tube diameter of: *a)* 11.0 Å and $(p / p_0)$ = 0.9, and 18.1 Å *b)* $(p / p_0)$ = 0.01 and *c)* $(p / p_0)$ = 0.9. Concentration increases from dark blue to red. The solid black lines indicate the centre of the nanotube walls. In the bottom graph ($D$ = 18.1 Å), the color scale for interstitial adsorption is 12 times larger than that for intratubular adsorption. Note the absence of interstitial adsorption at $D$ = 11.0 Å, and the apparent hollow cylindrical volume at $D$ = 18.1 Å, contained inside the fluid central layer and with centre at $(x, y)$ = (0 , 0).

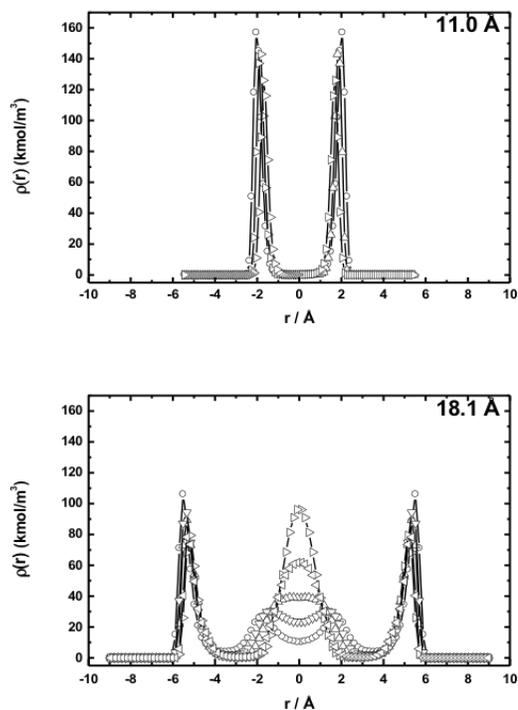

**Figure 6. Radial concentration profiles on the intratubular volume of a bundle at $(p / p_0)$ = 0.9.** Results for C$_2$H$_4$ and C$_3$H$_6$ at 11.0 Å are not plotted for simplicity sake, although the results are analogous to C$_2$H$_6$ and C$_3$H$_8$ (see text for details). ○ CH$_4$, △ C$_2$H$_6$, ▽ C$_2$H$_4$, ▷ C$_3$H$_8$, ◁ C$_3$H$_6$.

The intratubular concentration field gradient recorded in Figure 5c represents a real situation where two phases coexist inside the intratubular volume, corresponding to a monolayer of molecules adsorbed close to the tube walls, and to a second condensed phase on the central intratubular space. This concentration distribution has been observed for all fluids studied. We can define a radial concentration profile for the molecular centre of mass, $\rho(r)$, where $r$ is the distance to the nanotube central axis. The plots of $\rho(r)$ as a function of distance are recorded in Figure 6, for the smallest, 11.0 Å, and largest, 18.1 Å, nanotube diameters. The results for C$_2$H$_4$ and C$_3$H$_6$ at 11.0 Å, are very similar to their saturated analogues, so for simplicity sake they are not explicitly indicated in the graph. The peak intensities are related to the number of molecules that can be accommodated inside an annular volume of width $\Delta r$, so of course it decreases from methane to propane, *e.g.*, it is dependent on the number of carbon atoms on the molecule. Also, because unsaturated molecules are slightly smaller, the peak intensities also increase slightly when compared to the corresponding saturated analogue by less than 5 %.

It is worth noting the very different fluid structure observed in the two limiting situations. Whilst at 11.0 Å the molecules form only one annular adsorbed layer, arranged around the wall (Figure 5a and Figure 6), at 18.1 Å the intratubular volume is large enough to accommodate a second concentrical layer of fluid, whose structure is dependent on the carbon skeleton of the molecule. At these thermodynamical conditions of pressure and temperature, methane and ethylene can be supercritical, so we can not objectively discuss the existence of a liquid phase. However, one can observe that all fluids exhibit an annular layer of adsorbed molecules immediately around the tube inner walls, with density that decreases with the number of carbon atoms, as evidenced by the two main peaks centred at $r \approx -5.5$ Å and $r \approx 5.5$ Å (Figure 6). Beyond this initial layer, and as one moves towards the nanotube centre, fluids can also form a second intratubular phase, located around the nanotube centre ($r = 0$), and this layer structure now becomes dependent of the molecular identity. In the case of methane and ethylene, this second layer of confined molecules is discontinuous for it exhibits in its interior volume an hollow space of $\Delta r \approx 3.4$ Å (CH$_4$) and $\Delta r \approx 2.7$ Å (C$_2$H$_4$). This latter unoccupied volume is absent in the corresponding second layer formed by ethane, propane and propylene, which is a continuous one located inside $-3$ Å $< r < 3$ Å.

The densities plotted in Figure 6 are based on a nanotube volume defined by the skeletal diameter $D$ of the tube (Figure 1), *e.g.*, measured as the distance between centres of opposite carbon atoms on the walls. From the point of view of molecular simulation this definition is very convenient, but it is inconsistent with the accurate thermodynamical setting of the fluid-solid boundary as already pointed out before.[57, 58] Inspection of Figure 6 clearly indicates a non-negligible annular volume adjacent to the wall that is not accessible to fluid molecules; this exclusion volume arises from the strong short-ranged repulsive interactions between fluid and nanotube. A physically realistic definition for the thickness of that annular repulsive layer, is the largest distance from the solid wall to where the molecular centre of mass radial density is zero; using this definition we obtain $\delta$ ($CH_4$) ≈ 2.80 Å, $\delta$ ($C_2H_6$) ≈ 2.95 Å, $\delta$ ($C_2H_4$) ≈ 2.95 Å, $\delta$ ($C_3H_8$) ≈ 3.30 Å, $\delta$ ($C_3H_6$) ≈ 3.35 Å. Previous calculations at lower temperatures (210 K – 240 K) have determined $\delta$ ($C_3H_8$) ≈ 3.15 Å,[59] in very good agreement with our present result.

**B. Bundle exterior volume: grooves and rounded surface**

Similarly to what has been observed for molecular confinement onto the bundle interior volume, exterior adsorption also starts as a plateau on the isosteric heat curves (Figure 2), persistent until $(p / p_0) \approx 10 \times 10^{-3}$, then decreases until reaching a second (smaller) plateau after which it proceeds decreasing until fluid condensation. The general decreasing behavior of $q_{st}$ with monolayer coverage is according to what has been observed before, in experimental adsorption measurements of ethane onto bundles of closed-ended SWCNTs.[60] Tubes with small diameters do not exhibit interstitial adsorption, so they essentially possess external surface and groove sites available to physisorb molecules. Our curves seems to highlight the fact that there exists two different kinds of adsorptive sites available for fluid molecules, as evidenced by the two different plateaus mentioned before. Calbi *et al* performed detailed energy calculations on the external volume of an homogeneous bundle, and concluded that groove sites are energetically more favourable for methane adsorption than surface sites.[23] A similar conclusion has been obtained by Johnson and co-workers[24] who have shown that groove sites exhibit larger binding energies than surface sites**.** Therefore we attribute the first plateau to molecular adsorption onto groove sites, and the second one to adsorption onto surface sites. In Figure 4 we have plotted the ratios of molecular distribution between groove sites, $N^{(g,v)}$, and total number of adsortive sites, $N^{(v)}$, both in the grooves and curved surface; by definition, when $N^{(g,v)} = N^{(v)}$ we obtain $N^{(g,v)} / N^{(v)} = 0.5$. Therefore when $N^{(g,v)} / N^{(v)} > 0.5$ most molecules will be adsorbed onto groove sites. A closer look at Figure 4 seems to indicate that in the low pressure region there is threshold of $D = 13.8$ Å above which molecules tend to be more adsorbed onto groove sites. As nanotube diameter increases above the previous 13.8 Å threshold, so does the predominance of groove adsorption, until reaching a limiting value of $N^{(g,v)} / N^{(v)} \approx 0.89$ for propane at a diameter of $D = 18.1$ Å. When pressure starts to increase, and grooves become totally occupied, adsorption onto the outer rounded surface will be the predominant mechanism and thus $N^{(g,v)} / N^{(v)}$ will start decreasing. Cruz *et al* have observed that propane and propylene adsorption onto an homogenous bundle of 18.1 Å diameter tubes, starts to occur on groove sites between two adjacent tubes.[33] Whilst groove occupation is weakly dependent on individual tube diameter, adsorption of molecules onto surface sites is rather independent from that geometrical constraint. This is not surprising if one bears in mind the local geometry of a bundle exterior volume: the dimensions of a groove site are more diameter-dependent than the corresponding surface sites adjacent to that same groove. In fact, previous calculations of methane adsorption onto groove and surface sites have pointed out that for a small molecule as methane, the binding energy on the bundle exterior volume was independent from individual tube diameter.[24] In the present study this finding is corroborated for methane but also verified for the other chain molecules.

The second plateau on the $q_{st}$ curves is more pronounced for propane and propylene, highlighting the ability of those molecules to better adapt themselves on the rounded outer surface of a nanotube; this reflects the possibility of enhanced molecular bending in the $C_3$ species. A similar argument holds when comparing molecules with the same molecular weight, but with unsaturated chemical bonds. The inhibition of internal twisting degrees of freedom on ethylene and propylene, caused by the existence of a rigid π bond, results in these molecules having slightly smaller isosteric heats than ethane and propane, respectively. Nonetheless, we have determined $q_{st}$ average values for molecules adsorbed onto surface sites, obtained from the second plateau of the isosteric heat curves recorded in Figure 3: $q_{st}^S$ ($CH_4$) = 12.48 ± 0.05 kJ/mol, $q_{st}^S$ ($C_2H_6$) = 16.45 ± 0.05 kJ/mol, $q_{st}^S$ ($C_2H_4$) = 15.13 ± 0.05 kJ/mol, $q_{st}^S$ ($C_3H_8$) = 21.27 ± 0.14 kJ/mol and $q_{st}^S$ ($C_3H_6$) = 20.59 ± 0.02 kJ/mol. The corresponding results are plotted in Figure, and, as can be observed, isosteric heat values can be approximated by a linear function of diameter with angular slope ≈ 0.

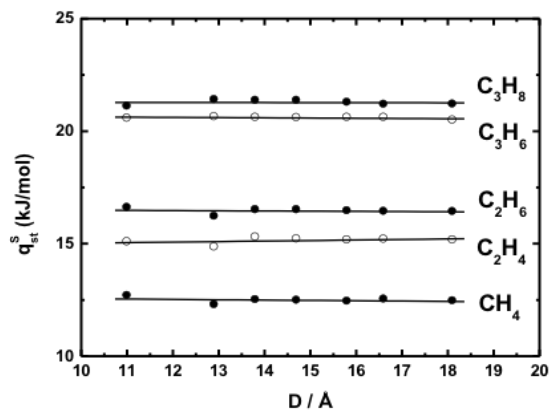

**Figure 7.** Isosteric heats of adsorption on the bundle's external rounded surface as function of individual tube diameter. Note the constancy of all calculated data, whose average errors are smaller than the symbols size.

As mentioned in the previous section, methane and ethylene can exhibit supercritical behaviour at the highest relative pressures studied here. However, for the subcritical fluids (ethane, propane and propylene), we can extrapolate the exterior adsorption $q_{st}$ curves towards $(p / p_0) = 1$ in order to estimate their enthalpies of vaporization. For all the individual nanotube diameters we employed a least-squares method to correlate data in the range $(p / p_0) > 0.5$, whose extrapolation towards fluid condensation lead to the following results: $\Delta H_{vap}$ ($C_2H_6$) = 13.4 ± 0.1 kJ/mol, $\Delta H_{vap}$ ($C_3H_8$) = 16.3 ± 0.1 kJ/mol, and $\Delta H_{vap}$ ($C_3H_6$) = 16.0 ± 0.1 kJ/mol. Note the relatively very small standard deviations of the results, indicating that enthalpies of vaporization are independent from tube diameter. The obtained results are thoroughly consistent with experimental data,[61] 9.76 kJ/mol ($C_2H_6$), 16.25 kJ/mol ($C_3H_8$) and 16.04 kJ/mol ($C_3H_6$).

### C. Zero-coverage isosteric heats of adsorption

Under high vacuum conditions, or zero-coverage region, the fluid-fluid contribution to the isosteric heat of adsorption can be safely neglected, thus allowing one to obtain further insight into the nature of the fluid-solid interactions. Data recorded in Figure 3 have been extrapolated towards the limit $p \to 0$, allowing the calculation of zero-coverage isosteric heats, $q_{iso}^0$. The corresponding results for confinement on the bundle's interior and exterior volume are shown in Figure 8. As mentioned before and even at very low bulk pressures, the lighter fluids (methane, ethane and ethylene) can become adsorbed onto interstitial channels whose single tube diameter surpasses 15.8 Å ($CH_4$), 16.6 Å ($C_2H_4$) and 18.1 ($C_2H_6$). This is particularly relevant for the larger diameter, where low pressure interstitial adsorption can account for 14 % – 29 % of total bundle interior adsorption, but rather unimportant for the smaller diameters where it never exceeds 1 % – 5 % of total adsorption. Bearing this in mind, zero-coverage data indicated in Figure 8 excludes interstitial adsorption for fluids adsorbed onto bundles of 18.1 Å individual tube diameter. Although not explicitly indicated in Figure 8, we have also calculated $q_{iso}^0$ at 18.1 Å considering the possibility of interstitial adsorption and it deviated *ca.* 18 % ($CH_4$), 8 % ($C_2H_4$) and 13 % ($C_2H_4$) from pure intratubular adsorption. A general trend can be established, namely that $q_{iso}^0$ scales with molecular weight. As molecular weight increases from $C_1$ to $C_3$, so does the number of interacting sites on the molecule (pseudoatoms), thus enhancing the fluid interactions with the solid wall.

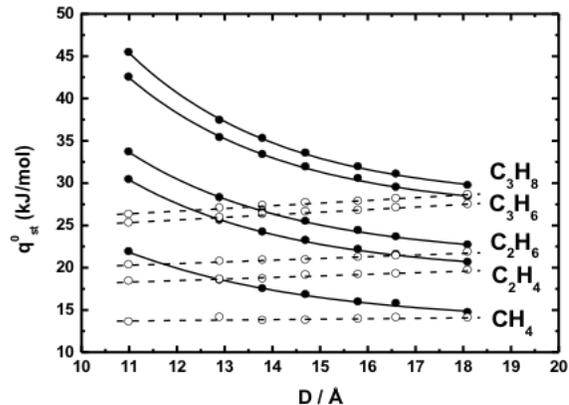

**Figure 8. Zero-coverage isosteric heats of adsorption on the bundle's interior (filled symbols) and exterior volume (open symbols).** Calculated values at 18.1 Å for $CH_4$, $C_2H_6$ and $C_2H_4$ do not consider interstitial confinement (see text for details). Symbols are calculated data and lines are either linear (dashed) or first order exponential (solid) decays (eqn. 5).

It is evident that $q_{st}^0$ slowly increases linearly for adsorption onto the bundle's external volume, according to $q_{st}^0 = A \cdot D + \left(q_{st}^0\right)^0$, but with increasing slope moving from $C_1$ to $C_3$ (open symbols of Figure 8). When internal adsorption is to be considered, the zero-coverage isosteric heat decreases according to the first order exponential decay of eqn. 5 (closed symbols of Figure 8). A closer inspection suggests that there is a crossover point where external adsorption lines will eventually cross the corresponding internal ones, when the former replace the latter as the most energetically favorable site for adsorption. That crossover seems to be located in the range $D = 18 - 19$ Å. This fact is particularly interesting, and has been previously overlooked. After 19 Å, one can indeed expect two things to happen: *i)* the tube diameter reaches such a large value, that the bundle's external surface no longer behaves as a corrugated surface, but instead as a discontinuous sequence of very high mountain peaks (rounded surface) and very low valleys (grooves), and at the same time *ii)* the intratubular cross section becomes so large, that molecules are no longer able to feel the attraction of the wall immediately opposite to the one where they lye adsorbed, thus behaving as if they were confined in a graphite slit pore of very high spacing between the walls. We have performed a least-squares fitting for both dependencies, in the external and internal volume, and the resulting parameters are recorded in Table III.

$$q_{st}^0 = A \cdot \exp\left(\frac{-D}{B}\right) + \left(q_{st}^0\right)^0 \qquad (5)$$

For comparison purposes, we have also calculated the zero-coverage isosteric heats of adsorption onto a single graphene layer, $q_{st}^{Gr}$ (1-layer), and onto an infinite number of stacked layers, $q_{st}^{Gr}$ (∞-layers), as indicated in Table III. The interaction potential between a Lennard-Jones site of an adsorbate particle and a single graphene sheet was given by Steele's 10–4 potential:[48]

$$U_{sf}(z) = 2\pi\rho_s\varepsilon_{sf}\sigma_{sf}^2 \left[\frac{2}{5}\left(\frac{\sigma_{sf}}{z}\right)^{-10} - \left(\frac{\sigma_{sf}}{z}\right)^{-4}\right] \quad (6)$$

where $z$ is the shortest distance between the Lennard-Jones adsorbate site and the graphene layer, $\rho_s$ = 0.382 Å$^2$ is the atomic surface density of the SWCNT wall, and $\varepsilon_{sf}$ and $\sigma_{sf}$ are the solid-fluid cross parameters. If the Lennard-Jones adsorbate site interacts with an infinite number of stacked graphene sheets, with interlayer spacing $\Delta_s$ = 3.4 Å, then the potential energy of interaction can be calculated according to Steele's 10–4–3 equation:

$$U_{sf}(z) = 2\pi\rho_s\varepsilon_{sf}\sigma_{sf}^2 \left[\frac{2}{5}\left(\frac{\sigma_{sf}}{z}\right)^{-10} - \left(\frac{\sigma_{sf}}{z}\right)^{-4} - \frac{\sigma_{sf}^4}{3\Delta_s(0.61\Delta_s + z)^3}\right] \quad (7)$$

**Table III.** Zero-coverage isosteric heat of adsorption dependency with individual SWCNT diameter[a] and partial contributions from intratubular, $q_{st}^{0,I}$, and interstitial, $q_{st}^{0,IC}$ adsorption calculated at 18.1 Å. For comparison purposes, experimental isosteric heats of adsorption on a graphite basal plane, $q_{st}^{Gr}$, are also indicated, along with adsorption onto a single graphene layer, $q_{st}^{Gr}$ (1-layer), and an infinite number of stacked layers, $q_{st}^{Gr}$ (∞-layers).

|  |  | (kJ/mol) | CH$_4$ | C$_2$H$_6$ | C$_2$H$_4$ | C$_3$H$_8$ | C$_3$H$_6$ |
|---|---|---|---|---|---|---|---|
| Interior | | $A$ | 141 ± 39 | 303.4 ± 0 | 268 ± 34 | 538 ± 62 | 446 ± 67 |
| | | $B$ (Å) | 3.9 ± 0.4 | 3.4 ± 0.0 | 3.4 ± 0.2 | 3.3 ± 0.0 | 3.3 ± 0.0 |
| | | $(q_{st}^0)^0$ | 13.5 ± 0.4 | 21.1 ± 0.0 | 19 ± 0.0 | 26.3 ± 0.0 | 26.6 ± 0.3 |
| | | $(q_{st}^{0,I})^b$ | 14.7 ± 0.0 | 22.6 ± 0.0 | 20.6 ± 0.0 | 33.1 ± 0.3 | |
| | | $(q_{st}^{0,IC})^b$ | 24.1 ± 0.0 | 35.2 ± 0.1 | | | |
| Exterior | | $A$ (Å$^{-1}$) | 0.05 ± 0.03 | 0.20 ± 0.01 | 0.18 ± 0.02 | 0.33 ± 0.01 | |
| | | $(q_{st}^0)^0$ | 13.1 ± 0.4 | 18.1 ± 0.2 | 16 ± 0.0 | 20.9 ± 0.0 | 20.1 ± 0.2 |
| | | $q_{st}^{Gr}$ | 14.9[29,c] | 16.0 – 19.7[62,d] | | 24.8 – 27.3[d] | |
| | | $q_{st}^{Gr}$ (1-layer) | 9.13 ± 0.07 | 14.64 ± 0.09 | 13.23 ± 0.05 | 19.20 ± 0.09 | 18.39 ± 0.07 |
| | | $q_{st}^{Gr}$ (∞-layers) | 10.85 ± 0.09 | 17.37 ± 0.09 | 15.70 ± 0.07 | 22.74 ± 0.08 | 21.68 ± 0.06 |

[a] Parameters for interior adsorption were correlated with eqn. 5 and for exterior adsorption with a linear function of diameter, $q_{st}^0 = A \cdot D + (q_{st}^0)^0$. [b] Data for $D$ = 18.1 Å in the range 8 × 10$^{-7}$ < ($p$ / $p_0$) < 9 × 10$^{-3}$ have been used. [c] Determined at 77 K. [d] Determined at 300 K.

As far as internal adsorption is concerned and as individual tube diameter increases, eqn. 5 tends towards an asymptotic behavior whose limit is in fact $(q_{st}^0)^0$. For methane, $(q_{st}^0)^0$ = 13.5 kJ/mol is remarkably similar to adsorption onto the basal plane of graphite, confirming our previous arguments that there is a finite and limiting diameter after which intratubularly confined molecules behave as if they were adsorbed onto a slit-like pore. A similar observation holds for external adsorption, but this time when nanotube diameter becomes so narrow, that the consequence on the bundle's external surface is a flattening towards a planar graphene sheet. In this last case, we obtained for methane $(q_{st}^0)^0$ = 13.1 kJ/mol, in reasonable agreement with the experimental value of 14.9 kJ/mol for methane adsorption onto the basal plane of graphite[29] and with our own calculated value for adsorption onto a infinite number of stacked graphene layers of 10.85 kJ/mol. The comparison of our results for C$_2$H$_6$ and C$_3$H$_8$ exterior and interior adsorption, indicate that in both cases our calculated data are in satisfactory agreement with the corresponding experimental quantities.[62] Jiang et al have calculated the zero-coverage isosteric heats of methane, ethane and propane on homogeneous open-ended bundles with a single tube diameter of 13.56 Å.[32] Under those conditions, interstitial adsorption is inhibited and their reported values can be considered as resulting from intratubular confinement. Their data, $q_{st}^0(CH_4)$ = 18.27 kJ/mol, $q_{st}^0(C_2H_6)$ = 27.71 kJ/mol, and $q_{st}^0(C_3H_8)$ = 35.86 kJ/mol compares remarkably well with ours, calculated at a diameter of 13.8 Å, $q_{st}^0(CH_4)$ = 17.49 kJ/mol, $q_{st}^0(C_2H_6)$ = 26.66 kJ/mol, and $q_{st}^0(C_3H_8)$ = 35.25 kJ/mol.

It should be noted that for the largest nanotube studied in the present work, corresponding to $D$ = 18.1 Å, intertubular spacing is large enough to accommodate methane, ethane and ethylene molecules in the interstitial channels at very low pressures. Any relevant propane and propylene interstitial confinement only occurs at relatively large pressures, typically ($p$ / $p_0$) ≥ 0.01 (cf. Figure 4). At that diameter, it becomes possible to separate the individual contributions from intratubular and interstitial adsorption. In fact, the calculated zero-coverage isosteric heat from interstitial adsorption is larger than the corresponding quantity for intratubular confinement by ca. 60%, a discrepancy which is rather independent from molecular length or chemical nature (Table III). At such large tube diameters, molecules at the interstitial channels experience adsorption on a nanovolume that is confined between three different nanotube walls, enhancing the solid-fluid interactions. On the other hand, molecules inside the tube energetically feel as if they were confined in a slit-pore geometry, of large spacing between the two planar walls, and thus subject to weaker solid-fluid interactions. Therefore, due to energetic considerations, it should be expected that the latter is less favorable for adsorption than the former.

It is known that at low pressure, adsorption on the bundle exterior volume starts in the groove sites, and proceeds to surface sites as chemical potential increases.[23, 29] Therefore, our results of $(q_{st}^0)^0$ for external adsorption are a good measure of zero-coverage isosteric heats on the groove sites. An interesting outcome is the fact that interstitial sites exhibit larger isosteric heats than groove sites in the bundle with larger diameter (*cf.* Table III). Indeed, energy calculations of methane adsorption onto closed-ended heterogeneous bundles, point out to the fact that in the case of large diameter tubes the corresponding interstitial sites have equal or larger isosteric heats than the groove sites.[24, 28] We plotted the values of $(q_{st}^0)^0$ recorded in Table III, as a function of the number of carbon atoms on the molecule, $N_C$, in Figure 9. The result is very interesting but not totally unexpected, in the sense that a linear trend is obtained for adsorption both on the bundles exterior and interior volume. That interior adsorption (intratubular and interstitial) limiting properties are linear with the number of carbon atoms, has already been observed before for a bundle with an individual tube diameter of 13.56 Å.[32] The present results not only corroborate those findings, but also extend them for the process of exterior adsorption and for unsaturated molecules. From the inspection of Figure 9, one can observe that $(q_{st}^0)^0$ for $CH_4$ is rather insensitive to either exterior or interior confinement, but this similarity no longer holds when the fluid molecule becomes longer, and both curves start to diverge away from each other. The trend for interior adsorption indicates a higher sensitivity towards the number of carbon atoms on the molecule, as evidenced by the corresponding curve larger slope.

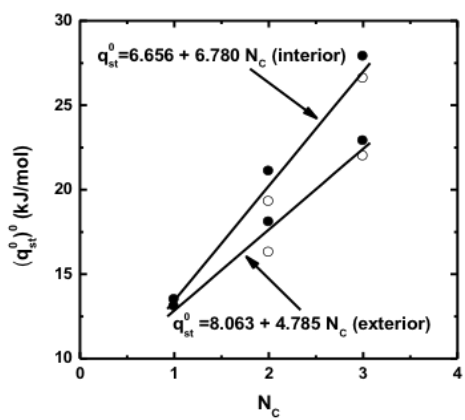

**Figure 9. Limiting adsorption zero-coverage isosteric heats on the bundle's interior and exterior volume**, corresponding to the parameters $(q_{st}^0)^0$ recorded in Table III. $CH_4$, $C_2H_6$ and $C_3H_8$ (filled symbols), $C_2H_4$ and $C_3H_6$ (open symbols).

## IV. CONCLUSIONS

Although our bundle geometrical model is an approximation to a realistically prepared sample, we were able to access the contributions from the four different adsorptive sites on the bundle and establish general trends. Interstitial adsorption is only significative at the largest diameter employed (18.1 Å), but at that diameter it is energetically more favorable than intratubular confinement. Moreover, from the plots of intratubular radial concentration profiles, we were able to estimate a physically realistic distance of closest approach, $\delta$, of a fluid molecule towards the solid wall; this distance decreases with the number of carbon atoms on the fluid molecule. These results enable the definition of an accurate thermodynamical pore effective volume as $D_{eff} = D - \delta$, where D is the skeletal diameter of the nanotube, measured between centres of opposite carbon atoms on the walls. Moreover, we have verified that in the low pressure region adsorption onto groove sites is the dominant mechanism, and that adsorption onto surface sites is rather independent from individual tube diameter.

The results obtained for the zero-coverage isosteric heats produced a significant improvement of the established understanding of the solid-fluid interactions. The corresponding calculated data for intrabundle confinement, are very dependent on individual tube diameter, but that dependency tends to be smeared out as diameter increases and reaches an asymptotic behavior. Confinement on the bundle interior volume is always energetically more favorable than exterior adsorption, in the whole pressure range, and that difference increases with pressure and molecular weight. Differences in the zero-coverage isosteric heat between saturated and unsaturated molecules are small, 1 – 3 kJ/mol, and the unsaturated compound exhibits the smaller value; its molecular skeleton is more rigid and therefore less able to adapt to an adsorptive site. However, a previously unaccounted threshold seems to exist at $D = 18 – 19$ Å, after which external adsorption can become dominant over intrabundle confinement. In the present study, low-pressure interstitial adsorption is only relevant for methane and the $C_2$ components at an individual tube diameter of 18.1 Å. In those cases, it is possible to establish a relative order of zero-coverage isosteric heats as, $(q_{st}^0)^{interstitial} > (q_{st}^0)^{intratubular} > (q_{st}^0)^{grooves} > (q_{st}^0)^{surface}$.

### ACKNOWLEDGEMENTS

F.J.A.L. Cruz gratefully acknowledges financial support from *F.C.T./M.C.E.S.* (PT) through grant SFRH/BPD/45064/2008 and J.P.B. Mota acknowledges financial support from *F.L.A.D.* (Portugal).